\documentclass[aip,graphicx]{revtex4-1}

\usepackage{graphicx}
\usepackage{epstopdf, epsfig}
\usepackage{subfig}
\usepackage{amsmath}
\usepackage{color}
\usepackage{graphicx}
\usepackage{epstopdf, epsfig}
\usepackage{multirow}
\usepackage{array}

\begin{document}

\title{Entrainment and mixing in gravity currents using simultaneous velocity-density measurements}

\author{Sridhar Balasubramanian}
\affiliation{Department of Mechanical Engineering, IIT Bombay, Mumbai, MH 400076, India}
\affiliation{Interdisciplinary Program in Climate Studies, IIT Bombay, MH 400076, India}
\affiliation{ 
Corresponding author: sridharb@iitb.ac.in 
}

\author{Qiang Zhong}%
\affiliation{College of Water Resources $\&$ Civil Engineering, China Agricultural University, Beijing 100083, China}
\affiliation{Beijing Engineering Research Center of Safety and Energy Saving Technology for Water Supply Network System, Beijing 100083, China}

\date{\today}

\begin{abstract}
Gravity currents modify their flow characteristics by entraining ambient fluid, which depends on a variety of governing parameters such as the initial density, $\Delta \rho$, the total initial height of the fluid, $H$, and the slope of the terrain, $\alpha$, from where it is released. Depending on these parameters, the gravity current may be designated as sub-critical, critical, or super-critical. It is imperative to study the entrainment dynamics of a gravity current in order to have a clear understanding of mixing transitions that govern the flow physics, the shear layer thickness, $\delta_{u}$, and the mixing layer thickness, $\delta_{\rho}$. Experiments were conducted in a lock-exchange facility in which the dense fluid was separated from the ambient lighter fluid using a gate. As the gate is released instantaneously, an energy conserving gravity current is formed, for which the only governing parameter is the Reynolds number defined as $Re=\frac{Uh}{\nu}$, where $U$ is the front velocity of the gravity current, and $h$ is the height of the current. In our study, the bulk Richardson number, $Ri_{b}$=$\frac{g^{'}H}{U_{b}^{2}}$=1, takes a constant value for all the experiments, with $U_{b}$ being the bulk velocity of the layer defined as $U_{b}$=$\sqrt{g^{'}H}$. Simultaneous Particle Image Velocimetry (PIV) and Planar Laser Induced Fluorescence (PLIF) measurement techniques are employed to get the velocity and density statistics. A flux-based method is used to calculate the entrainment coefficient, E$_{F}$, for a Reynolds number range of $Re\approx$400-13000 used in our experiments. The result shows a mixing transition at $Re\approx$2700 that is attributed to the flow transitioning from weak Holmboe waves to Kelvin-Helmholtz type instabilities. Following this mixing transition, the entrainment coefficient continued to increase with increasing Reynolds number owing to the occurrence of Kelvin-Helmholtz billows that promote small scale local mixing and cause another marked spike in the flux entrainment values. The results confirmed a non-monotonic nature of scalar mixing in lock-exchange gravity currents. Experimentally, it was also observed that the flux entrainment value near the front of gravity current was 2-9 times higher than the head value depending on the value of the Reynolds numbers.

\end{abstract}


\maketitle

\section{Introduction}

Gravity currents or ``density/turbidity currents" are common in environmental and engineering flows, and they are driven by horizontal pressure gradients arising due to density variations. Some examples are the katabatic winds, ocean overflows, sea-breeze fronts, thunderstorm and microburst outflows, avalanches, and hydrothermal vents. The important parameters determining gravity currents propagation are the density difference between the two fluids ($\Delta\rho$), gravity ($g$), the depth of the gravity current ($h$), total depth of the fluid layer ($H$), and the slope of the terrain ($\alpha$). Owing to the Boussinesq approximation, $g$ and $\Delta\rho$ can be combined to form the buoyancy jump or reduced gravity, $g^{'}=\frac{g\Delta\rho}{\rho_{0}}$, where $\rho_{0}$ is the reference density $\rho_{0}=\frac{\rho_{1}+\rho_{2}}{2}$. For the simple case of a dam-break generated gravity current on a horizontal bottom ($\alpha$=0), wherein a barrier separating the dense and the lighter fluids of depth $H$ is removed to generate the current, dimensional considerations show that $h$ is related to $H$. For miscible fluids, entrainment and mixing at the edges and within the current play a crucial role in the structure and dynamics of the gravity current \cite{ET, Linden1979, Huppert1980, EJS1982, Fernando, Manins, Princevac2005}.

The mixing between fluids of different densities due to entrainment is a continuing research problem with a long history \cite{Eckart1948, MTT, Hunt1983, Turner1986, Strang2001a, Strang2001b, Harish2015}. Different forms of entrainment have been identified \cite{Hunt1983}, the most common being entrainment into a mean current of characteristic velocity $U$ through a flow with velocity $w_{H}$ normal to the interface, which is directly application to gravity currents (Figure 1). The Morton-Turner-Taylor (MTT) entrainment hypothesis was developed for this case \cite{Turner1986}. For a gravity current propagating horizontally on a surface,  $U\propto$$\sqrt{g^{'}H}$ is taken to be the front velocity of the gravity current.

The active regions of gravity currents have been well established \cite{EJS1982, Simpson1979b}. The interface between two fluids close to the head of a gravity current is a typical frontal zone, that is, a region in which, notwithstanding intense mixing, a high density gradient is present. The frontal zone is immediately followed by the head, which has some fractional depth of the initial height, $H$, depending on the nature of the gravity current. The scaling for velocity and depth of two counter-flowing gravity currents, produced by lock-exchange, has been well understood \cite{EJS1982, Shin2004, Cantero2007}. For the Boussinesq case, \cite{Yih1965} proposed that the depths of two currents are equal in height, $h$=$\frac{H}{2}$, along their entire lengths. The speed of both gravity currents are the same and have the value proposed by\cite{Ben1968} for energy-conserving gravity currents. \citet{Klemp1994} argued, based on shallow-water theory that idealized energy-conserving gravity currents cannot be realized in a lock-exchange initial-value configuration, as the speed of this current would be faster than the fastest characteristic speed in the channel predicted by the shallow-water theory. Extensive measurements show that on a horizontal surface the characteristic front velocity of a gravity current is given by $U$ = 1.05$\sqrt{g^{'}h}$, where $h$ is the depth of the steady current \cite{Keulegan1957, Ben1968}. These results were mainly obtained from flows occupying about 1/5 of the total depth $H$, but recent work with lock-exchange flows has shown that $U$ is sensitive to changes in the value of $\frac{h}{H}$ in the range 1/3 to 1/10 $H$ as proposed by \citet{Simpson1979b}.They argue that the inviscid gravity current depth can never be greater than 0.3473$H$, wherein, according to Benjamin's theory \cite{Ben1968}, the gravity current has its fastest speed. Therefore, Benjamin's theory for energy conserving gravity currents is widely accepted and the velocity and depth of a gravity current is given as,

\begin{equation}
U=0.4\sqrt{g^{'}H}, \quad h=\frac{H}{2}
\label{Eq:1}
\end{equation}

Most previous studies have focussed on the dynamics of the head, turbulence dissipation and mixing, as well as scaling for the front velocity and fractional depth. \citet{ET} quantified gravity current entrainment as a function of bulk Richardson number, $Ri_{b}=\frac{g^{'}H}{U_{b}^{2}}$. In their configuration the bulk Richardson number was variable, since the inertial and buoyancy forces were decoupled. In the present case, however, the governing parameters are g' and H, and in view of \ref{Eq:1} the bulk Richardson number is a constant. For this genre, the only possible variable is the Reynolds number $Re$=$\frac{Uh}{\nu}$=$\frac{UH}{2\nu}$ \cite{Simpson1979b}, which has been consistently used for lock-exchange flows \cite{Shin2004, Cantero2007}.

The entrainment coefficient can be mathematically defined as 

\begin{figure}
\centerline{\includegraphics[width=10cm]{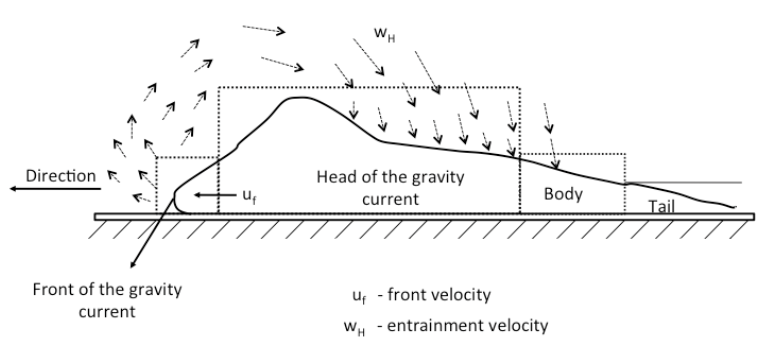}}
\caption{Profile of a self-adjusting gravity current, similar to that proposed by \citet{EJS1982}.}
\label{fig:1}
\end{figure}

\begin{equation}
E=\frac{w_{H}}{U_{c}}
\label{Eq:2}
\end{equation}
which for homogeneous (non-density-stratified) environments is a constant that depends on the type of flow. \citet{Hunt1983} proposed different forms of entrainment coefficients, for example, based on the volume conservation,

\begin{equation}
w_{H}=\frac{\partial(Uh)}{\partial x},
\label{Eq:3}
\end{equation}
the thickness of the mixing layer, $\delta$,

\begin{equation}
w_{B}=\frac{d\delta}{dt}
\label{Eq:4}
\end{equation}
which is the boundary entrainment rate, and the fluxes across the interface,

\begin{equation}
w_{F}=\frac{\rho^{'}w^{'}}{\Delta\rho}
\label{Eq:5}
\end{equation}
which is the flux entrainment velocity. The form $E(Ri)$ is known as the 'entrainment law', which has been studied extensively using laboratory and field experiments as well as numerical simulations \cite{Fernando, Princevac2005}.  For gravity currents configuration of \citet{ET}, the following form was proposed,

\begin{equation}
E=\frac{0.08-0.1Ri}{1+5Ri}, \quad Ri\leq0.8
\label{Eq:6}
\end{equation}
Conversely, for self-adjusting currents of the form discussed in this paper,  $Ri$=constant$\approx$1, and hence E is only a function of Re.
%
%
%
%
%
%
%
%
%
%
\begin{equation}
E=\Phi(Re)
\label{Eq:9}
\end{equation}
In fact, Ri is a constant for such energy conserving gravity currents \cite{Cantero2007}. 

In the earlier works on turbulent entrainment \cite{Eckart1948}, three phases of turbulent mixing have been identified, namely, the initial, middle, and final stages. In the initial stage, the mixing is mostly driven by large-scale flow structures that control turbulent mixing. In the middle stage, intermediate scales control turbulence and mixing along with large scales. In the final stage, small scale mixing occurs, possibly determined by local instabilities that transition to three-dimensional turbulence. As a practical matter, the latter requires a sufficiently high Reynolds number \cite{Batchelor59,Hinze1975}. Although with some care laminar state can be maintained with increasing Reynolds numbers up to suitably high Reynolds numbers, the converse is not true: turbulence cannot be sustained if the Reynolds number falls below some minimum value.

In the present work, we study the effects of Reynolds number on entrainment into a gravity current produced by lock-exchange, focusing on different regimes of turbulent mixing. The focus is on the dependence on flux entrainment coefficient, $E_f=\frac{w_f}{U}$ as a function of Reynolds number of the gravity current. The characterization of shear layer thickness, $\delta_{u}$, and the scalar mixed layer thickness, $\delta_{P}$ as a function of Re is also undertaken. This characterization is only done for the front and head of the gravity current, since the entrainment is vigorous in these zones.

\section{Theoretical Background}

Consider a collapsing gravity current produced by a lock-exchange mechanism. The density of the dense fluid is $\rho_{2}$, the lighter fluid is $\rho_{1}$, and the release height is $H$. The stream-wise and normal coordinates, $x$ and $z$, respectively are fixed, and the mean flow is two-dimensional. A schematic of such a current is given in Figure \ref{fig:1}, which is characterized by the front, head, body, and tail (the definitions are given later). The region below the depth $h <\frac{H}{3}$ is tagged as the body of the gravity current followed by its tail (for which $h<\frac{H}{6}$). Our interest here is on studying the entrainment dynamics of the front and head of a gravity current. The flux entrainment can be found from the conservation of buoyancy, details of which follows.

For a 2-D infinitesimal element (assuming per unit width), for fluxes in the stream-wise ($x$) and vertical ($z$) directions, the governing equation for buoyancy is

\begin{equation}
\frac{\partial b}{\partial t}+\frac{\partial (bu)}{\partial x}+\frac{\partial (bw)}{\partial z}=0
\label{Eq:10i}
\end{equation}
where $b$=$\overline{b}$+$b^{'}$ is the total buoyancy defined as $b$=$\frac{g(\rho-\rho_{r})}{\rho_{r}}$, where $\rho_{r}$ is some reference density. The buoyancy flux term, $b$$u_{i}$, in $x$ and $z$ directions include the contributions from both the mean (e.g. $\overline{b} U$) and fluctuating (e.g. $b^{'} u^{'}$) components. Utilizing the idea of Reynolds averaging, one could re-write \ref{Eq:10i} as follows,

\begin{eqnarray*}
\frac{\partial \overline{b}}{\partial t}+{U_{j}}\frac{\partial\overline{b}}{\partial x_{j}}=\frac{\partial q_{x}}{\partial x}+\frac{\partial q_{z}}{\partial z}\\
\end{eqnarray*}
where $q_{i}$=-$\overline{b^{'}u^{'}_{i}}$ is the buoyancy flux from the fluctuating part of the flow. It should be noted in the above equation that the mean convection term is primarily due to the stream-wise component ($U$), in which the gravity current propagates. The mean normal velocity is very small and can be approximated to $W\approx$0. Thus the above equation reduces to 

\begin{equation}
\frac{\partial \overline{b}}{\partial t}+U\frac{\partial\overline{b}}{\partial x}=\frac{\partial q_{x}}{\partial x}+\frac{\partial q_{z}}{\partial z}
\label{Eq:11i}
\end{equation}
\\
For the above equations, if a Lagrangian approach is used, where a moving frame of reference is considered, and the infinitesimal element moves with the front velocity (U) of the current, then equation \ref{Eq:11i} reduces to
\begin{equation}
\frac{\partial\overline{b}}{\partial t}=\frac{\partial q_{x}}{\partial x}+\frac{\partial q_{z}}{\partial z}
\label{Eq:12i}
\end{equation}

An uncertainty arises while applying equation \ref{Eq:12i} at the head of the gravity current due to the difference in propagation velocities of the front and the head of the current. This gives rise to a mean velocity difference, $\delta U$=$U-U_{h}$, and hence some unaccounted fluxes. Here, $U$ is the velocity of the front of the current and $U_{h}$ is the velocity at the head of the current (such that $U_{h}<U$, since the head lags the front). We argue that the fluxes arising due to this difference, $\Delta U$, are negligible in comparison to the turbulent fluxes that play an important role in entrainment. The Lagrangian equation at the head of the current, assuming $U$ to be the reference frame, we arrive at,

\begin{equation}
\frac{\partial\overline{b}}{\partial t}+\delta U\frac{\partial\overline{b}}{\partial x_{j}}=\frac{\partial q_{x}}{\partial x}+\frac{\partial q_{z}}{\partial z}
\label{Eq:11ii}
\end{equation}
Doing an order of magnitude scaling analysis, it can be shown that the convection term ($2^{nd}$ term) in the above equation has an order $\frac{\delta U}{u^{'}}$. From the available experimental measurements on mean velocity, it was seen that this ratio was small such that its order was $O<1$ $(\delta U< <  u^{'})$. For all the experimental runs, the value of velocity fluctuation, $u^{'}$, was found to an order of magnitude higher than $\delta U$. In lieu of this, the effect of the convective terms could be neglected in the Lagrangian frame of reference and the front velocity $U$ can be used as the relative frame while applying equation equation \ref{Eq:12i} to both the front and head of the gravity current. 
\\
Now, integrating equation (10) in the stream-wise direction we get,
\begin{equation}
\int_{0}^{L}\frac{\partial\overline{b}}{\partial t}dx=\int_{0}^{L}(\frac{\partial q_{x}}{\partial x}+\frac{\partial q_{z}}{\partial z})dx
\label{Eq:13i}
\end{equation}

\begin{equation}
\int_{0}^{L}\frac{\partial\overline{b}}{\partial t}dx=(q_x)_L-(q_x)_0+\int_{0}^{L}(\frac{\partial q_{z}}{\partial z})dx
\label{Eq:13a}
\end{equation}
The above equation can be rewritten as 



\begin{eqnarray}
\frac{\partial}{\partial t}\int_{0}^{L}\overline{b}dx=(q_x)_L-(q_x)_0+\frac{\partial}{\partial z}\int_{0}^{L}q_{z}dx\\
\frac{\partial\overline{b}^x}{\partial t}=\frac{1}{L}[(q_x)_L-(q_x)_0]+\frac{\partial\overline{q_z}^{x}}{\partial z}
\label{Eq:13b}
\end{eqnarray}
where $\overline{b}^{x}$=$\frac{1}{L}\int_{0}^{L}\overline{b}dx$ is the stream-wise averaged value of mean buoyancy and the term $\overline{q_z}^{x}$=$\frac{1}{L}\int_{0}^{L}{q_z}dx$ is the stream-wise averaged vertical buoyancy flux with $L$ being the length of the control volume chosen. Integrating equation (15) in the vertical direction ($z$) gives,

\begin{equation}
\int_{0}^{H}\frac{\partial\overline{b}^{x}}{\partial t}dz=\{\int_{0}^{H}\frac{1}{L}[(q_x)_L-(q_x)_0] dz\}+(\overline{q_z}^{x})_H-(\overline{q_z}^{x})_0
\label{Eq:14}
\end{equation}
At the bottom, $z$=0, the buoyancy flux should become zero, owing to the wall boundary condition. Hence equation \ref{Eq:14} reduces to

\begin{equation}
\int_{0}^{H}\frac{\partial\overline{b}^{x}}{\partial t}dz=\{\int_{0}^{H}\frac{1}{L}[(q_x)_L-(q_x)_0] dz\}+(\overline{q_z}^{x})_H
\label{Eq:14a}
\end{equation}
Equation \ref{Eq:14a} summarizes that the depth integrated rate of change of stream-wise averaged mean buoyancy flux is equal to the sum of depth integrated change in the horizontal flux and the stream-wise averaged vertical flux at the top boundary. The horizontal buoyancy flux, $q_x$, is generally smaller compared to the vertical flux, $q_z$, owing to the strong interfacial gradients present in the vertical direction. Furthermore, the difference in the depth integrated horizontal fluxes between the inlet and outlet would be even smaller. Therefore, the rate of change of buoyancy flux, to first order, can be approximated to the vertical flux entering or leaving the system at the top boundary. Under this assumption \ref{Eq:14a} can be reduced to 

\begin{equation}
\int_{0}^{H}\frac{\partial\overline{b}^x}{\partial t}dz=(\overline{q_z}^{x})_H
\label{Eq:14c}
\end{equation}
\\
The region of interest is equally divided into small control volumes and a stream-wise averaged value of the mean buoyancy $\overline{b}^{x}$ is calculated. The flux entrainment velocity is then found from \ref{Eq:14c} by vertically integrating the time rate of stream-wise averaged mean buoyancy present in the gravity current, i.e.,

\begin{equation}
w_{F}=\frac{1}{g}\int_{0}^{H}\frac{\partial\overline{b}^{x}}{\partial t}dz
\end{equation}
\\
This definition of $w_F$ is similar to that given in equation \ref{Eq:5}. Using this method, a direct estimate of the flux based entrainment coefficient is obtained near the front and the head of the gravity current. The body and tail of the current are not considered in this study owing to weak entrainment dynamics compared to the front and the head. Once the flux entrainment velocity, $w_F$ is calculated using the above method, the flux entrainment coefficient is obtained as

\begin{equation}
E_{F}=\frac{w_{F}}{U}
\label{Eq:15}
\end{equation}
where $U$ is front velocity of the current. All the entrainment results presented below are calculated using Equation \ref{Eq:15}.

\section{Experimental Facility and Measurement Techniques}

Below we present the details of the experimental facility used for generating an energy conserving gravity current. This is followed by explanation on laser optical based measurement technique used to measure the velocity and density fields in the flow.

\subsection{Lock-Exchange Facility}
\begin{figure}
\centerline{\includegraphics[width=8cm]{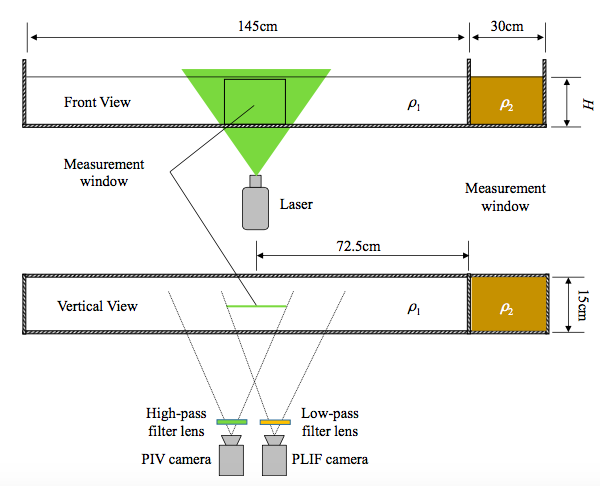}}
\caption{Schematic of the experimental set up and measurement techniques.}
\label{fig:2a}
\end{figure}

The experiments were conducted in a plexiglass tank equipped with a lock-exchange mechanism. The tank dimensions were 175 cm long, 15 cm wide and 30 cm high. A schematic of the apparatus is shown in Figure \ref{fig:2a}. The tank was separated into two parts by a lock gate located at 30 cm from right end. The dense fluid $\rho_{2}$ in the right slot occupies a predetermined depth $H$ before the gate is released. The dense solution is prepared by adding requisite amount of NaCl to water and mixing it to get a uniform density fluid. The rest of the tank is filled with lighter ambient fluid, $\rho_{1}$, to the same depth $H$, which is separated by the gate. Upon removing the gate instantly, a gravity current is initiated due the difference in the hydrostatic pressure between the two fluids. The quick motion of the gate ensures that perturbations due to gate opening are extremely small, and no relative fluid motion is generated in the direction of the pull. Thus, there are is no secondary flows or disturbances due to the gate release. The denser gravity current undercuts the lighter fluid, which flows in the opposite direction. The measurement section is located at the middle of the tank. A continuous-wave 2W 532nm laser was used to create a 1-mm-thick light sheet to illuminate the middle vertical plane of the tank. The light sheet is projected from the tank bottom. High resolution Particle Image Velocimetry (PIV) and Planar Laser Induced Fluorescence (PLIF) cameras were set up at the same side of the tank to avoid interference. A high- and low-pass filter lenses were used for PIV and PLIF camera, respectively. The cutoff wavelength for both lens are 550 nm. 

The Reynolds numbers ranged from $Re$=485 to 12270, which cover flow transitioning from weak to strong mixing. The dense and light fluid were created using salt solution and an aqueous solution of ethanol, respectively. This salt-ethanol technique was introduced to match the refractive indices accurately, enabling the use of optical measurement techniques. This method ensured that the images quality is high, which allows accurate measurements. A densitometer (make: Mettle Toledo Densito 30PX) and a refractometer (make: Leica handheld analog refractometer) were used to match the refractive indices and measure the density of the two fluid. Details of the method of matching the refractive indices using salt and alcohol are given in \citet{Han1988}, \citet{Strang2001a}, and \citet{Xu2012}.

\subsection{Velocity and Density Measurements}
\begin{table}
\begin{center}
\def~{\hphantom{0}}
\begin{tabular}{lcccc}
  Case  &  Image size (pixel)   &  Frequency (fps) &   Resolution (pixel/mm)  &  Iterations\\[3pt] 
      C1 & 1312x432 & 50 & 9.25 & 2 \\														
      C2 & 1216x432 & 50 & 9.24 & 2 \\	    												
      C3 & 1200x400 & 50 & 9.17 & 2 \\			 										
      C4 & 1296x720 & 50 & 9.30 & 2 \\												
      C5 & 1248x912 & 50 & 9.58 & 2\\
      C6 & 1312x1056 & 50 & 9.21 & 2\\
      C7 & 1200x880 & 50 & 9.18 & 2\\
      C8 & 1200x886 & 50 & 9.18 & 3\\
      C9 & 1200x1072 & 50 & 9.28 & 3\\
      C10 & 800x1056 & 75 & 9.32 & 2\\
      C11 & 800x1056 & 75 & 9.32 & 3\\ 					
\end{tabular}
\caption{PIV Parameters.}
\label{tab:1i}
\end{center}
\end{table}
A time-resolved PIV system was used for obtaining instantaneous velocity fields in the $x$-$z$ plane. Hollow glass beads, 20 $\mu$m in median diameter and specific gravity, $SG$=1.1, were used as tracers for both dense and light fluids. Particle images were captured by a IDS UI-3360CP-M/C USB 3.0 camera, with a resolution of 2048x1088 pixels, was used for capturing the images of flow evolution. A 50mm f/1.4 lens was used with the camera, and the aperture value was reduced to f/2 to get an appropriate depth of field and reduced aberration. A high-frequency pass filter with the cutoff wavelength 550 nm was used with PIV camera for filtering out the fluorescence from PLIF dyes. 

Major parameters used for Particle Image Velocimetry are listed in Table \ref{tab:1i}. The camera was operated under continuous sampling mode for all cases. Particle images are analyzed by using the iterative multi-grid image deformation method proposed in \citet{Scarano}. The window size in the final iterative step is 16 pixels x 16 pixels. Two or three iterative steps were taken for different cases to match the velocity, resolution and sampling frequency to ensure the particle moving distances between PIV image pairs meet the one-quarter rule \cite{Adrian1991}. Detailed description of the methodology used for PIV can be found in \citet{Zhong2015}.

Planar Laser-Induced Fluorescence (PLIF) was employed to obtain instantaneous density fields in the present experiments. Rhodamine 6G (R6G) is used as the fluorescent dye that is mixed uniformly with the light fluid.Both the PLIF and PIV systems share the 532nm laser as illumination. The peaks of absorption and emission spectrum of R6G are around 530nm and 550nm respectively. PLIF images are recorded by a IDS uEye UI-1220-C USB 2.0 camera, which has a CMOS sensor with 752x480 pixels. PIV and PLIF cameras are synchronized by software trigger. A 8.5mm f/1.5 lens was used with this camera. A low-pass filter lens with cutoff wavelength 550 nm was located in front of the lens for filtering out the green reflected light from PIV particles. The image gray value, namely the fluorescence intensity, is used to obtain the local R6G concentration. When the incident light is much lower than the saturation intensity, the intensity of the fluorescence is proportional to the dye concentration \cite{Crimaldi}. The relationship between gray value and R6G concentration at each pixel is calibrated using the same calibration conditions. Following the procedure discussed in \citet{Xu2012}, the calibration was done using 11 different solutions with concentrations varying from 0 to 100 $\mu$g$L^{-1}$ with 10 $\mu$g$L^{-1}$increment. One thousand images of test section were recorded for each step to obtain average gray value at each pixel. 

The local R6G concentration can be found from the local gray value, and the local R6G concentration has a linear relationship with the local density. When the dense fluid (Volume $V$ and density $\rho_{2}$) and a lighter fluid (Volume $xV$ and density $\rho_{1}$, R6G concentration $C_{1}$) are mixed uniformly, the density and the R6G concentration of the mixture is:

\begin{eqnarray*}
\rho=\frac{\rho_{1}xV+\rho_{2}V}{xV+V}=\frac{\rho_{1}x+\rho_{2}}{x+1}\\
C=\frac{C_{1}xV}{xV+V}=\frac{xC_{1}}{x+1}
\end{eqnarray*}
Thus, if the local R6G concentration C is known, the local density can be found using:
\begin{equation}
\rho=\rho_{2}-\frac{C}{C_{1}}(\rho_{2}-\rho_{1})
\label{eq:10i}
\end{equation}
As seen from this equation, if the local concentration, $C$ equals the known concentration of the lighter fluid, $C_{1}$, then the local density is same as that of the lighter fluid. This is the initial state, when both the fluids are separated by the gate. Upon the release of the gate, entrainment occurs causing change in the local density.

\section{Results and Discussion}

\begin{table}
\begin{center}
\def~{\hphantom{0}}
\begin{tabular}{lccccc}
  $\rho_{1}$ (kgm$^{-3}$)  &  $\rho_{2}$ (kgm$^{-3}$)   &  $U$ (ms$^{-1}$)  &   $H$ (m)   &   $Re=\frac{Uh}{\nu}$   &  $Ri=\frac{g^{'}H}{U_{b}^{2}}$\\[3pt] 
       0.9961 & 1.0009 & 0.020 & 0.05 & 485 & 1\\														
       0.9904 & 1.0099 & 0.040 & 0.05 & 985 & 1\\	    												
       0.9804 & 1.029 & 0.062 & 0.05 & 1560 & 1\\			 										
       0.9903 & 1.0117 & 0.052 & 0.08 & 2080 & 1\\												
       0.9901 & 1.0089 & 0.055 & 0.1 & 2735 & 1\\									
       0.9932 & 1.007 & 0.051 & 0.12 & 3070 & 1\\								
       0.9804 & 1.026 & 0.085 & 0.1 &  4270 & 1\\													
       0.9716 & 1.046 & 0.109 & 0.1 & 5480 & 1\\	 												
       0.9803 & 1.0281 & 0.107 & 0.15 & 8050 & 1\\											
       0.9714 & 1.0468 & 0.135 & 0.15 & 10150 & 1\\
       0.9712 & 1.0471 & 0.144 & 0.17 & 12270 & 1\\					
\end{tabular}
\caption{Experimental parameter range.}
\label{tab:1}
\end{center}
\end{table}

\begin{figure}
\centerline{\includegraphics[width=8cm]{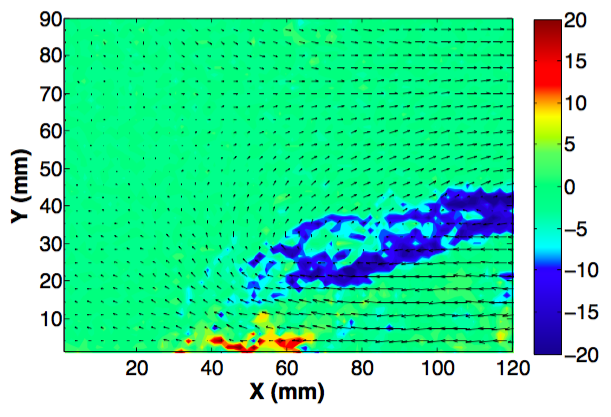}}
\caption{Vorticity profile near the front of the gravity current for Re=4270.}
\label{fig:2}
\end{figure}

A total of 11 experiments (see Table \ref{tab:1}) were conducted for varying values of Reynolds number, which was obtained by varying either the density difference, $\Delta\rho$, or the initial height of the dense fluid, $H$, or both. As already established before (in the introduction section), a gravity current that is initiated solely due to the potential energy of the fluid has a profile that can be divided into four regions:\\

1. \textbf{Front or Nose}: The front/nose is the region where the flux entrainment process begins, wherein, the ambient lower density fluid mixes with the propagating current via interfacial instabilities and boundary exchange. The front is usually associated with the highest gradients in the velocity and density. In addition to instabilities, strong vortex structures also cause mixing of the two fluids in the region, as shown in Figure \ref{fig:2}, where two counter rotating vortices can be seen. The front velocity of the gravity current is already defined in Section 1.\\

2. \textbf{Head}: Following the front, a gravity current consists of a characteristic ``head," that is deeper than the following flow. The head of the gravity current is associated with breaking waves and intense mixing process, and plays an important part in the behaviour of the current. As per \citet{EJS1982}, ``\textit{In a gravity current flowing horizontally this head remains quasi-steady, and is about twice as deep as the following flow, but in a current flowing down a slope the size of the head continually increases}''. The former is true for the present study, where the head remains steady throughout the flow of the gravity current. The flux entrainment is still active near the head region.\\

3. \textbf{Body}: The body is defined as the point when the fractional height drops below the critical value of 0.3$H$.  Here, the entrainment process reduced in intensity and the flux entrainment decreases.\\

4. \textbf{Tail}: It denotes the end of the gravity current and the region where the fractional height drops to 0.16$H$. This zone is associated with almost no or very little entrainment.\\

The front and the head of the gravity current, being the primary zones of mixing, and thus all the measurements in our experiments have been made in these two zones.\\
\begin{figure}
\centerline{\includegraphics[width=8cm]{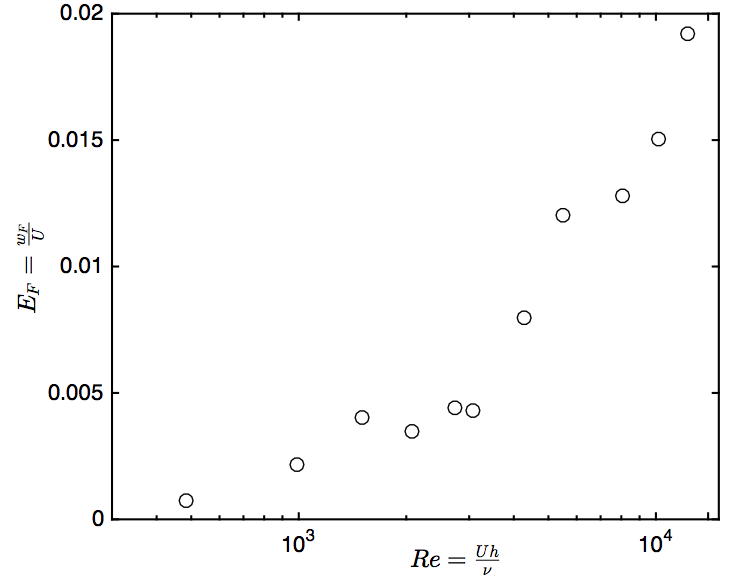}}
\caption{Averaged flux entrainment coefficient, $E_{F}=\frac{w_{F}}{U}$, near the head of the gravity current (as defined by Figure 1) as a function of Reynolds number. The mixing transition is clearly visible from the spikes in the values of $E_{F}$.}
\label{fig:3}
\end{figure}

The results from two liquid-phase shear layer measurements of \citet{Dimotakis1986} and \citet{Dimotakis2005}, depicts the estimated chemical product thickness as a function of the local Reynolds number. The definition of local Reynolds number uses the differences in the velocity between the shear layers ($\Delta U$) and the thickness, $\delta$ was defined as the 1$\%$ width of the total mixed-fluid probability P$_{m}$(y), where P$_{m}$ had dropped to 1$\%$ of its maximum value. Since this width was found to agree very well with the visual width $\delta_{vs}$ of the layer, the latter was used in their study. The mixing efficiency (entrainment) was measured as a ratio of scalar mixing thickness, $\delta_{P}$ to that of the visual width. This definition of entrainment is not a direct measure, but similar to equation (\ref{Eq:4}) defined above. The results from \citet{Dimotakis2005} indicated a marked increase in the mixing beyond Re$\approx$10$^{4}$, where the local definition of Reynolds number is used. Further increasing the Reynolds number above Re$\gg$10$^{4}$,  it was observed that the entrainment reached a self-similar value. It was also observed that, in variable density mixing, the scalar mixing actively modulates the flow dynamics, referred to as Level-2 mixing \cite{Dimotakis2005}. However, once the flow reaches a sufficiently high Reynolds number, the flow becomes fully turbulent and is controlled mostly by three-dimensional mixing. According to \citet{Dimotakis2005}, this transition occurs at Re$\approx$10$^{4}$. 

In general, there has been little consensus on the criteria for mixing transition in variable density flows. Furthermore, this criterion for transition could differ depending on the type of the flow \cite{Dimotakis2005}. The experiment by \citet{Dimotakis1986} provides limited information about mixing transition, since a direct measure of entrainment was not obtained in their study. This led to overestimation of the absolute amount of chemical product in their study. As reported by \citet{Hunt1983}, there are several methods used for calculation of entrainment coefficient, but the flux method is believed to be most accurate and exact. The mixing due to both large-scale coherent structures and small-scale swirling eddies is accounted for while using the flux method. The results based on such a method are described in the present study. The definition of Reynolds number used in our study is consistent with the local Reynolds number definition proposed by \citet{Dimotakis2005}. The mixing transition was quantified based on the flux entrainment values computed using equation \ref{Eq:15}.

In Figure \ref{fig:3}, the variation of flux entrainment coefficient, $E_{F}$=$\frac{w_{F}}{U}$, at the head of the gravity current, as a function of the Reynolds number is shown. Since the head of the current extends to a finite stream-wise distance, $x$, the value presented is the averaged over this distance. At the lowest Reynolds number ($Re$=485), the mixing and entrainment is very small as evidenced from the low value of flux entrainment coefficient. It can also be seen that the value of flux entrainment remains low up until $Re$=$\frac{Uh}{\nu}\approx$2700, after which there is a marked increase in the value of $E_{F}$. The reason for the low values of $E_{F}$ for Re$\approx$400-2700 is due to the fact that the mixing is primarily controlled by Holmboe waves. This mechanism has been documented in shear stratified flows by other researchers as well \cite{Thorpe1972, Strang2001a, Princevac2005, Ivey2003}. The Holmboe waves are generally associated with weak mixing, since they are interfacial waves and have very thin mixing layers. As we increase the Reynolds number beyond $Re>$2700, the flux entrainment, $E_{F}$ also increases, indicating a first mixing transition. Following the transition, the value of $E_{F}$ continues to increase with increasing Reynolds number up to $Re\approx$5500. In this region, the high level of mixing is due to the formation of Kelvin-Helmholtz (K-H) instabilities, which initiates localized mixing at the boundary of the gravity current in form of K-H vortex rolls (seldom referred to as primary K-H instability). \cite{Strang2001a}

\begin{figure}
\centerline{\includegraphics[width=8cm]{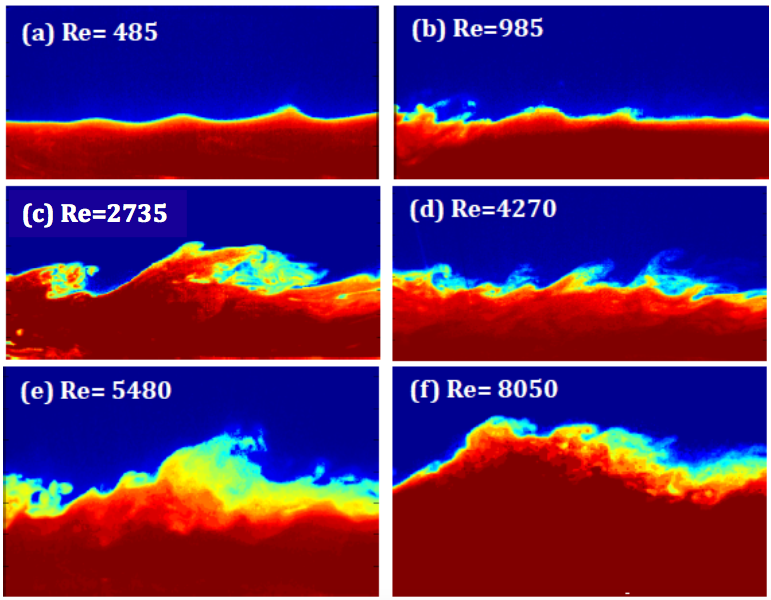}}
\caption{Density images qualitatively capturing the differences in the mixing at different Reynolds number.}
\label{fig:4}
\end{figure}

\begin{figure}
\centerline{\includegraphics[width=8cm]{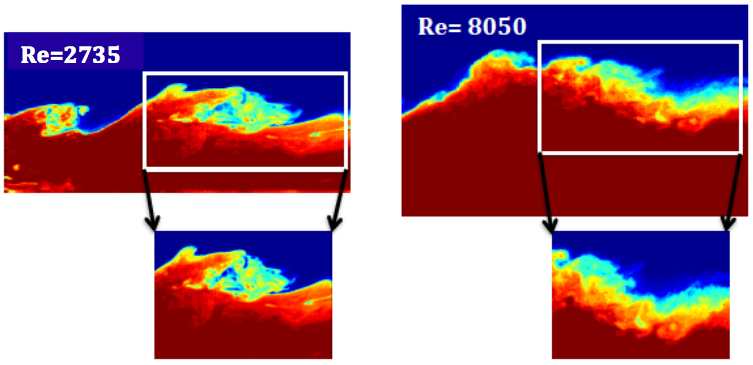}}
\caption{Zoom in view of the nature of instability driven mixing at low and high Reynolds numbers.}
\label{fig:5}
\end{figure}

Following the first mixing transition, upon further increasing the value of Reynolds number (Re$>$8000), a second transition is observed, where the value of flux entrainment, $E_{F}$, increases significantly and doesn't reach a self-similar value as document by \citet{Dimotakis2005}. The increase in the entrainment in this region is due to the breakdown of K-H rolls, which aid in the formation of Kelvin-Helmholtz (K-H) billows. These K-H billows initiate small- scale mixing and thereby transition the flow to a highly turbulent state characterized by three-dimensional (3-D) mixing. The billows, formed from the breakdown of the primary Kelvin Helmholtz vortex into small scale eddies, lead to an efficient mixing of the dense and the ambient fluid. This inturn causes an increase in the flux entrainment. Figure \ref{fig:4} shows the different regimes of mixing starting with Holmboe waves, followed by primary K-H vortex rolls, and finally the K-H billows. The qualitative nature of mixing is conspicuously seen in Figure \ref{fig:4}, which corroborates the flux entrainment trend shown in Figure \ref{fig:3}. A zoomed image of the mixing nature of the flow at two different Reynolds number is shown in Figure \ref{fig:5}. It is visible that at moderate Reynolds number of $Re$=2735, the mixing is primarily driven by Kelvin-Helmholtz rolls. At a higher Reynolds number, $Re$=8050, small scale mixing overrides large-scale flow structures leading to a higher mixing of the fluids. At very low Reynolds numbers (i.e., $Re\leq$2000, not shown in Figure \ref{fig:5}), Holmboe waves are dominant.

\begin{figure}
\centerline{\includegraphics[width=12cm]{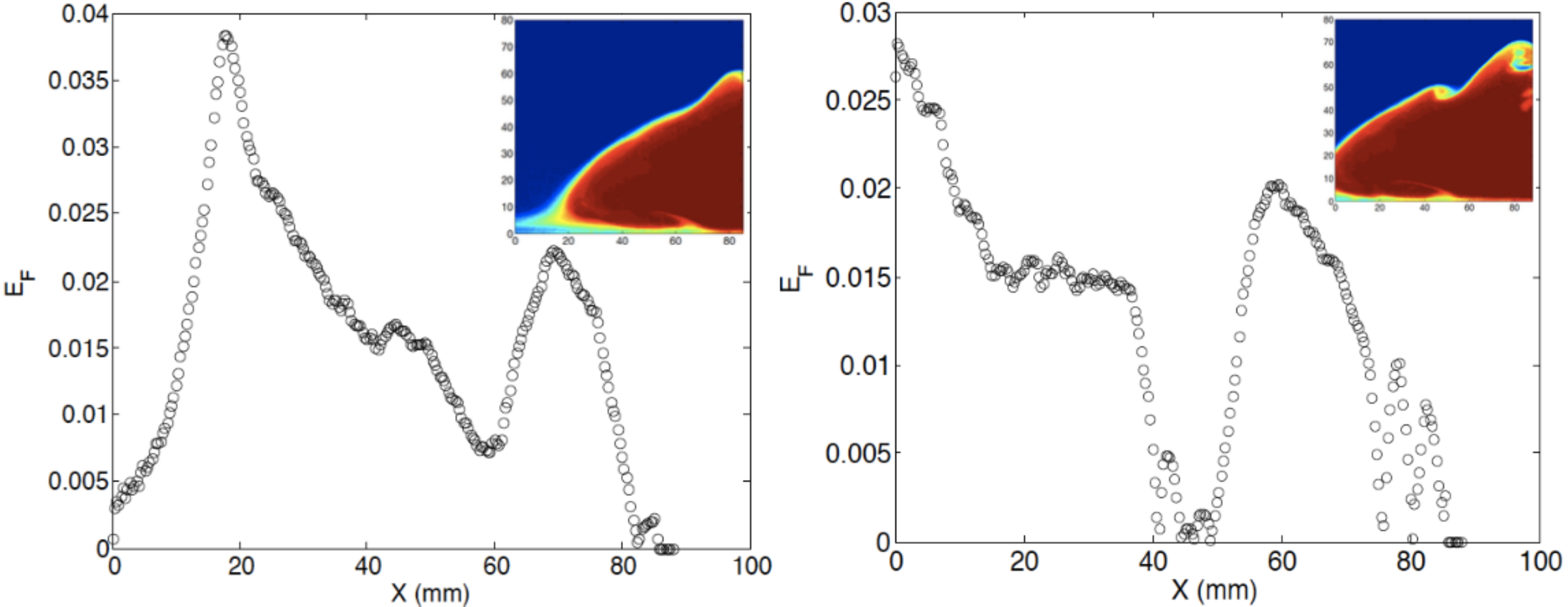}}
\caption{Comparison of flux entrainment coefficient, $E_{f}=\frac{w_{F}}{U_{b}}$, at the front of the gravity current and its head for $Re$=12270.}
\label{fig:6}
\end{figure}

In Figure \ref{fig:6}, the flux entrainment as a function of stream-wise distance ($x$) is shown at two different time instances. The image frame (see insets in Figure \ref{fig:6}) chosen is such that the front/nose of the gravity current starts at $x\approx$15 mm followed by the head. The distance $x\approx$0-10 mm primarily comprises of only the lighter fluid ($\rho_{1}$) as seen in Figure \ref{fig:6}a (left image), but the effect of entrainment on the propagating gravity current is also felt in this region, evidence from the very low value of $E_{F}$. At $x\approx$20 mm, the entrainment peaks indicating vigorous mixing near the front of the current. Following this peak, the value starts decreasing, thereby representing the passage of the head of the current. 

A slightly different picture emerges from Figure \ref{fig:6}b (right image). The current covers the entire stream-wise span in this figure. Hence, the value of entrainment starts from a high value and gradually decreases. From both the graphs and corresponding insets in Figure \ref{fig:6}, it is seen that the flux entrainment is high near the front of the current and in regions where the fluid rolls up, indicating strong mixing regions. This feature of entrainment was recently observed by \citet{Ottolenghi2016} in their numerical simulations. Another interesting feature of entrainment that was observed in our experiments was that the flux entrainment value near the front of the propagating gravity current was always higher than the value at the head of the gravity current; a fact corroborated from Figure \ref{fig:6}.

\begin{table}
\begin{center}
\def~{\hphantom{0}}
\begin{tabular}{lccc}
      $Re$    &  $A:E_{F}$ @ Front   &    $B:E_{F}@$ Head &   Ratio ($\approx$$\frac{A}{B}$)\\[3pt]
       485   & 0.0063 & 0.00073 & 9\\
       985   & 0.020 & 0.0022 & 8\\
       1560  & 0.035 & 0.004 & 8\\
       2080  & 0.026 & 0.0036 & 7\\
       2735 & 0.021 & 0.0044 & 7\\
       3070 & 0.030 & 0.0043 & 7\\
       4270 & 0.032 & 0.008 & 4\\
       5480 & 0.05 & 0.012 & 4\\
       8050  & 0.043 & 0.013 & 3\\
       10150 & 0.032 & 0.015 & 2\\
       12270 & 0.040 & 0.019 & 2\\
\end{tabular}
\caption{Values of flux entrainment, $E_{F}$, near the front and head of the gravity current, and their ratios for different values of Reynolds numbers.}
\label{tab:2}
\end{center}
\end{table}

The ratio of value of flux entrainment at the front to the value at the head for each Reynolds number used in our study is given in Table \ref{tab:2}. From this table, we see that at low Reynolds numbers the mixing intensity near the front of the gravity current is about 8-9 times higher than that near the head, while with the increase in the value of Reynolds number, the ratio of flux entrainment at the front to the head drops to as low as 2 (see the value for $Re$=10150 and $Re$=12270). The reason is due to the differences in the type of mixing mechanism near the front and the head of the gravity current. Near the front, the mixing is primarily due to combination of vortex driven mixing (Figure \ref{fig:2}) and fluid entrainment. On the other hand, at the head of the gravity current, the mixing is primarily due to entrainment of ambient fluid into the propagating gravity current fluid. As seen from the mixing transition curve in Figure \ref{fig:3}, the entrainment spikes at higher values of $Re$, which translates to increased mixing near the head of the gravity current. Therefore, the ratio of entrainment, between the front and the head, decreases at high value of Reynolds number, Nevertheless, it is important to note that the front always has a higher value of mixing compared to the head of the gravity current, but the ratio has a strong dependence on the Reynolds number and the mixing transition regime.

In our opinion, the mixing transition based on the flux entrainment values, which gives the direct entrainment values as opposed to the other definitions of entrainment discussed in Section 1.1, accurately depicts the entrainment dynamics occurring at the head of the gravity current. Such a mixing transition has not been documented in previous literature available on this topic. The mixing transition documented in \citet{Dimotakis1986}, for mixing in chemically reacting gases, was calculated based on the non-dimensional mixing layer thickness, $\Delta$, defined as the ratio of shear layer thickness, $\delta_{s}$ to the vorticity layer thickness $\delta$. This definition of $\Delta$ is very different compared to the actual definition of entrainment used in our study. It was reported in \citet{Dimotakis1986} that the value of $\Delta$ reaches a self-similar value at very high Reynolds numbers, indicating a fully developed turbulence. Our results on flux entrainment capture the transition regimes, but don't capture the self-similar value at high Reynolds number, indicating possibility of further small-scale eddy driven mixing. This could be because $\Delta$ is not a good measure for parameterizing entrainment in variable density flows, since it doesn't take into account mixing at intermediate and small scales \cite{Dimotakis2005}. The definition for Reynolds number used in past studies is similar to the one used in our present study for energy conserving gravity currents. \cite{Cantero2007,ET}.

\begin{figure}
\centering
\begin{tabular}{ccc}
\subfloat{\includegraphics[width=5cm]{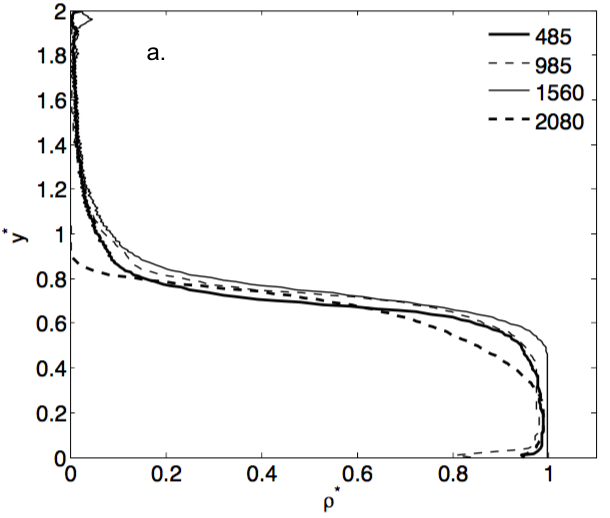}} & \subfloat{\includegraphics[width=5.1cm]{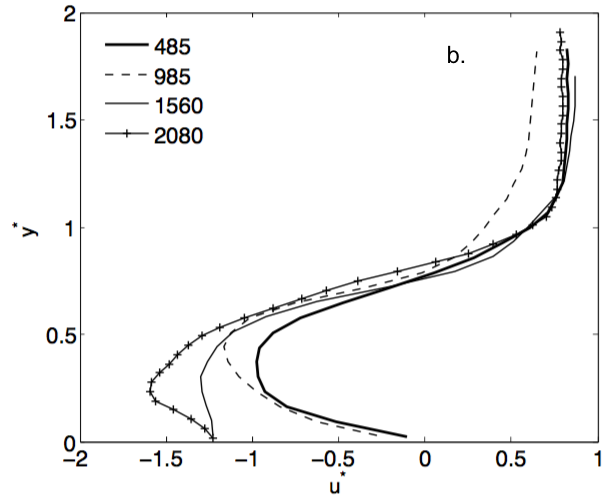}} \\
\subfloat{\includegraphics[width=5.1cm]{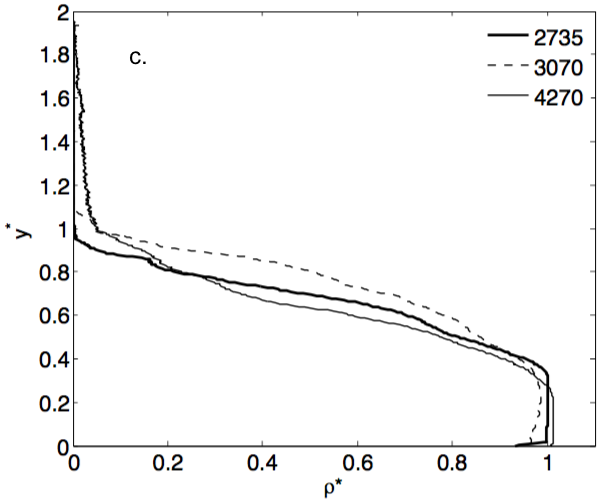}} & \subfloat{\includegraphics[width=5cm]{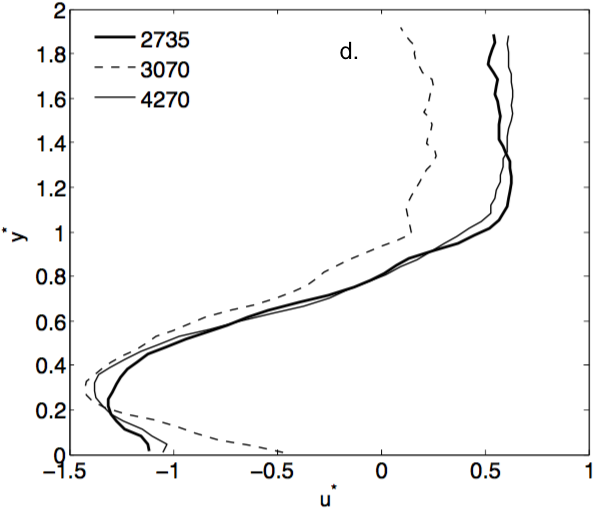}}\\
\subfloat{\includegraphics[width=5.1cm]{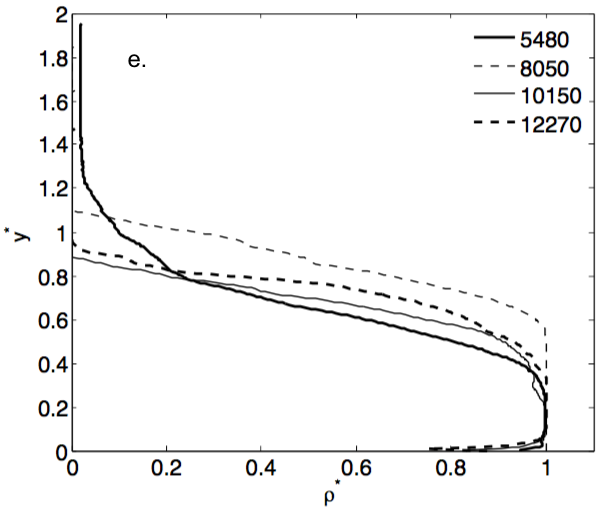}} & \subfloat{\includegraphics[width=5cm]{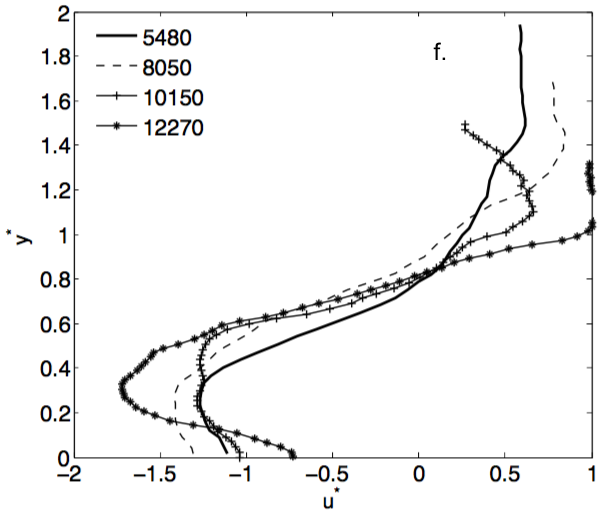}}\\
\end{tabular}
\caption{(a, c, e): Non-dimensional density profiles as a function of non-dimensional vertical height at low, moderate and high values of Reynolds numbers. (b, d, f) Non-dimensional velocity profiles as a function of non-dimensional vertical height at low, moderate and high values of Reynolds numbers.}
\label{fig:7}
\end{figure}

In order to further quantify our observations, profiles of velocity and density are plotted as a function of vertical distance for different values of Reynolds number. For a direct comparison, we non-dimensionalize the velocity ($u^{*}$) and density ($\rho^{*}$) profiles and their variation is shown against non-dimensional vertical height, $y^{*}$. From these profiles, values of velocity mixing layer thickness, $\delta_{u}$, and density mixing layer thickness, $\delta_{\rho}$, could be obtained, which gives better insight into the entrainment dynamics of a gravity current.

\begin{eqnarray*}
y^{*}&=&\frac{y}{h}\\
u^{*}&=&\frac{u}{U}\\
\rho^{*}&=&\frac{\rho-\rho_{1}}{\rho_{2}-\rho_{1}}
\end{eqnarray*}
Here $h$ is the height and $U$ is the front velocity of the gravity current. Briefly, $\delta_{u}$ is measured as the thickness in which the velocity shear is maximum. In a similar manner $\delta_{u}$ is measured as the thickness where the density variations are maximum. The non-dimensionalized density and velocity profiles are shown in Figure \ref{fig:7}a,b for four Reynolds number, namely, Re=455, 985, 1560, and 2080. For these values of Re, from Figure 4 and 5, we notice that the mixing is mainly due to the presence of weak Holmboe waves \cite{Smyth1991}. The study by \citet{Ivey2003} shows that for a Holmboe wave the velocity layer thickness is much higher, $\delta_{u}>$2$\delta_{\rho}$, which is clearly observed from Figure \ref{fig:7}a,b. Therefore, the amount of scalar mixing is restricted to a very thin layer.

The density mixing layer and the velocity mixing layer thicknesses for Re=2700, 3100, and 4200 are shown in Figure \ref{fig:7}c,d. For this range of Re, the strong Kelvin-Helmholtz rolls \cite{Smyth1991} dominate the mixing process, which has also qualitatively observed in Figure \ref{fig:4}. It is seen from this Figure \ref{fig:7}c,d that the density mixing layer thickness, $\delta_{\rho}$, increases and becomes comparable to the magnitude of velocity mixing layer thickness, $\delta_{u}$. Thus, the amount of mixing increases and the entrainment rate becomes higher leading to the first transition.

\begin{figure}
\centering
\begin{tabular}{ccc}
\subfloat{\includegraphics[width=5cm]{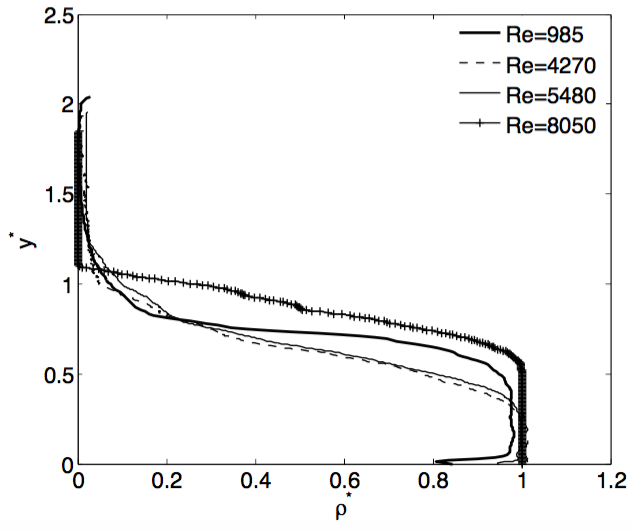}} & \subfloat{\includegraphics[width=5cm]{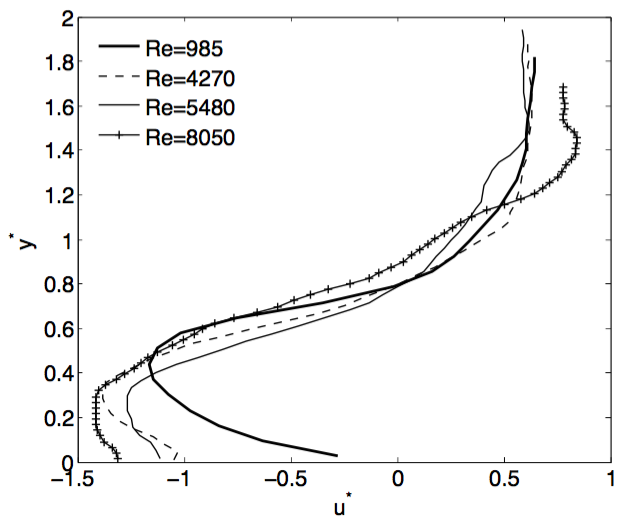}}\\
\end{tabular}
\caption{(Left): Non-dimensional density profile showing the change in the slope as the value of Reynolds number increases. (Right): Non-dimensional velocity profile showing invariance with increasing Reynolds number.}
\label{fig:8}
\end{figure}

The layer thicknesses for density and velocity for Re$>$5000 are shown in Figure \ref{fig:7}e,f. In this region, qualitatively, the mixing is dominated by Kelvin-Helmholtz billows (see Figure \ref{fig:4}). It is observed here that the density mixing layer thickness, $\delta_{\rho}$, further increases and equalizes with that of velocity layer thickness, $\delta_{u}$. The billows further increase the mixing between the two fluids, causing the second mixing transition shown in Figure 4.

Plotting the velocity and density mixing layer thickness for the transition Reynolds numbers, we see from Figure \ref{fig:8}, that as the flux entrainment increases, the slope of $\rho^{*}$, also increases. For the highest Reynolds number shown in this plot, we could see that the slopes of $u^{*}$, and $\rho^{*}$, are very similar, which is in line with the theory that as $\delta_{\rho}$ approaches $\delta_{u}$, the instability transitions from weak Holmboe to strong Kelvin Helmholtz type instability.

In order to summarize the different mixing transition zones, the values of the ratio of velocity and density mixing layers, defined as $\Delta^{*}=\frac{\delta_{u}}{\delta{\rho}}$ is shown in Table \ref{tab:3}, with an error estimate of $\pm0.1$mm. It is very clear from Table \ref{tab:3} that, for Holmboe waves the density based mixing layer is very small compared to the velocity mixing layer, and thus leading to lower values of entrainment. As we approach the first mixing transition at Re$\approx$2x10$^{3}$ and the other one at Re$\approx$8x10$^{3}$, one can observe the density mixing layer values increasing and thus the ratio, $\Delta^{*}$ approaches closer $\Delta^{*}$=1.
\begin{table}
\begin{center}
\def~{\hphantom{0}}
\begin{tabular}{lcccc}
 Re &  $\delta_{u}\pm0.1$ (mm)  &  $\delta_{\rho}\pm0.1$ (mm)  &   $\Delta^{*}$=$\frac{\delta_{u}}{\delta{\rho}}$ & Regime\\
       485 & 15.65 & 6.54 & 2.39 & Holmboe waves\\														
       985 & 15.54 & 7.35 & 2.12 & Holmboe waves\\	    												
       1560 & 16.84 & 8.01 & 2.1 & Holmboe waves\\			 										
       2080 & 16.65 & 8.42 & 1.97 & Holmboe waves\\												
       2735 & 31.71 & 29.87 & 1.06 & Kelvin Helmholtz vortex\\		
       3070& 39.85 & 35.87 & 1.11 & Kelvin Helmholtz vortex\\	
       4270 & 36.32 & 35.23 & 1.03 & Kelvin Helmholtz vortex\\	
       5480 & 39.58 & 36.49 & 1.08 & Kelvin Helmholtz billows\\	
       8050 & 40.42 & 39.20 & 1.03 & Kelvin Helmholtz billows\\	
       10150 & 43.08 & 42.18 & 1.02 & Kelvin Helmholtz billows\\	
       12270 & 46.308 & 46.215 & 1.002 & Kelvin Helmholtz billows\\											
\end{tabular}
\caption{Experimental parameter range.}
\label{tab:3}
\end{center}
\end{table}

A scaling law has been proposed for these two mixing layer thicknesses as follows \cite{Ivey2003},

\begin{equation}
\delta_{u}\approx C_{1}\frac{(\nu L)^{1/2}}{(g^{'}H)^{1/4}}
\label{Eq:11}
\end{equation}

\begin{equation}
\delta_{\rho}\approx C_{2}\frac{\kappa^{1/3}\nu^{1/6}L^{1/2}}{(g^{'}H)^{1/4}}
\label{Eq:12}
\end{equation}
where $C_{1}$ and $C_{2}$ are scaling coefficients, $\nu$ is the viscosity, $L$ is the horizontal length for which the interfaces are in contact, and $\kappa$ is the molecular diffusivity. The scaling coefficients were given a value $C_{1}$=$C_{2}$=1.5 based on the experimental results by \citet{Ivey2003}. Using the scaling laws, for exchange flows, one can come up with an estimate for bulk Richardson number, $Ri_{b}$, which is given as follows \cite{Ivey2003}.

\begin{equation}
Ri_{b}\approx 1.5\frac{(\nu L)^{1/2}}{(g^{'}H^{5})^{1/4}}
\label{Eq:13}
\end{equation}


The change in the depth, $H$, has a strong variation on the bulk Richardson number. This scaling is useful, since the classical formula for $Ri_{b}$, always gives $Ri_{b}$=1 for lock-exchange flows. From the scaling we can see that, as we increase the depth, the bulk Richardson number decreases indicating inertial flow with strong generation of instabilities. According to Equation \ref{Eq:13}, we report that the onset for transition from Holmboe instability to Kelvin-Helmholtz occurs at a Reynolds number $Re\approx$2000, or at a scaled bulk Richardson number $Ri_{b}\approx$0.073, which is in close agreement with the value of $J$=0.08, reported by \citet{Ivey2003} in their experiments. It should be noted that $J$ represents the bulk Richardson number as defined in \citet{Ivey2003}. Therefore, it could be concluded that the transition from mixing due to weak Holmboe waves to mixing governed by Kelvin-Helmhotz vortex rolls is expected to occur at a scaled bulk Richardson number of $Ri_{b}\approx$0.075. These scaling laws only give an approximate estimates of the Richardson number and hence for energy conserving gravity currents, it is better to relate entrainment as a function of Reynolds number for more generality.

\section{Conclusions}

The entrainment dynamics of an energy conserving gravity current, for which the bulk Richardson is $Ri_{b}$=1, was experimentally studied by measuring the velocity-density fields simultaneously. Based on the measurements, a direct method for calculating entrainment coefficient was adopted. This was termed as the \textit{``flux entrainment coefficient}'', $E_{F}$, and is defined as the ratio of flux-based entrainment velocity, $w_{F}$, to the front velocity, $U$, of the current. The flux entrainment method being a direct measure of the amount of mix between the fluids of varying densities is an important measure in such complex stratified flows. A control volume technique in a Lagrangian frame of reference was used to calculate $E_{F}$ at the front and the head of a propagating gravity current. A lock-exchange facility was used to produce gravity current of varying intensities. The entrainment in a lock exchange type horizontally propagating gravity current is only a function of the local Reynolds number, $Re$, defined as $Re$=$\frac{Uh}{\nu}$, which was varied between $Re\approx$400-13000 in our experiments.

The flux entrainment values were found to be the highest near the front of the gravity current due to vortex generated mixing coupled with entrainment exchange near the interface. Close to the head of the gravity current, the mixing was primarily due to the fluid exchange between the interfaces. The magnitude of entrainment coefficient between the front and the head of the gravity current was found to be higher at low Reynolds number indicating weak entrainment at those values. The flux entrainment coefficient as a function of Reynolds number showed three different mixing regions that were governed by the local flow dynamics. Two prominent mixing transition were captured and were quantified on the basis of the velocity layer and density layer thicknesses, $\delta_{u}$, and $\delta_{\rho}$ respectively. For Re$<$2700, the flux entrainment coefficient has a low value, and remains constant. In this region, the mixing is governed by weak Holmboe waves that propagate near the surface and don't create a strong mixing region. This region is quantified by a very thin density mixing layer, such that, $\delta_{u}>$2$\delta_{\rho}$, which was observed in our measurements. 

Following the Holmboe wave regime, increasing the value of Reynolds number above Re$\approx$2700, a mixing transition is observed characterized by higher values of flux entrainment coefficient, attributable to formation of Kelvin-Helmholtz vortex rolls. The K-H vortex perturbs the interface and causes vigorous mixing that penetrates deeper into the gravity current leading to higher entrainment of the ambient fluid into the current. In this region, the velocity layer thickness has the same magnitude as the density layer thickness, i.e., $\delta_{u}\approx\delta_{\rho}$ (i.e. $\Delta^{*}\approx$1). For Re$>$5000, the entrainment coefficient further increases due to the formation of Kelvin-Helmholtz billows, which cause local small scale mixing, thereby increasing the entrainment between the ambient fluid and the dense gravity current fluid. It was observed that the entrainment coefficient plateau at high Reynolds number due to vigorous mixing driven by K-H billows, which may have a complex three-dimensional structure. In this region as well $\Delta^{*}\approx$1. From these results, it could be concluded that the entrainment dynamics in a lock-exchange type gravity current undergoes mixing transition, which is strongly correlated to the type of instabilities causing entrainment and mixing of two fluids. We would like to state that from the present experimental regime, a transition to a fully turbulent state (i.e. a plateau in the valueof $E_{F}$) couldn't be captured, and experiments at higher Reynolds numbers (beyond Re=13000) need to be performed to reach a fully developed turbulent state. Experimentally, it was observed that the entrainment flux at the front of the gravity current is always greater than near the head for all values of Reynolds number, $Re$. The $E_{F}$ at the front is $\approx$ 8-9 times higher than the head value at low values of $Re$, and $\approx$ 2-4 times higher as the $Re$ value increases. This indicates vigorous scalar entrainment near the front and moderate entrainment at the head of the gravity current. From this, it could be concluded that the scalar entrainment near the body and tail would be very low. Therefore, the entrainment dynamics at the front and head of the gravity current play an important role in the scalar mixing of the current. The results from this study are meant to complement the existing literature on gravity currents and help in formulating a physics-based empirical parameterization for predicting entrainment dynamics in variable density flows.

\section*{Acknowledgements}
The experiments were conducted at University of Notre Dame, USA in the Department of Civil and Environmental Engineering and Earth Sciences. The authors are grateful to Prof. HJS Fernando for giving access to the experimental facility and technical discussions related to this work. This work was done when the first author (SB) was in University of Notre Dame as a Visiting Professor and the second author (QZ) was a Postdoctoral Fellow there during the visit period.

\bibliographystyle{jfm}
\bibliography{jfm-instructions}

\begin{thebibliography}{40}
\expandafter\ifx\csname natexlab\endcsname\relax\def\natexlab#1{#1}\fi
\def\au#1{#1} \def\ed#1{#1} \def\yr#1{#1}\def\at#1{#1}\def\jt#1{\textit{#1}}
  \def\bt#1{#1}\def\bvol#1{\textbf{#1}} \def\vol#1{#1} \def\pg#1{#1}
  \def\publ#1{#1}\def\arxiv#1{#1}\def\org#1{#1}\def\st#1{\textit{#1}}

\bibitem[Adrian(1991)]{Adrian1991}
{\sc \au{Adrian, R.J.}} \yr{1991}  \at{Particle-imaging techniques for
  experimental fluid mechanics}.  \jt{Annual Review of Fluid Mechanics}
  \bvol{23}~(1),  \pg{261--304}.

\bibitem[Baines(2002)]{Baines}
{\sc \au{Baines, P.G.}} \yr{2002}  \at{Two-dimensional plumes in stratified
  environments}.  \jt{Journal of Fluid Mechanics}  \bvol{471},  \pg{315--337}.

\bibitem[Batchelor(1971)]{Batchelor59}
{\sc \au{Batchelor, G.~K.}} \yr{1971}  \at{Small-scale variation of convected
  quantities like temperature in turbulent fluid. part 1. general discussion
  and the case of small conductivity.}  \jt{J.~Fluid Mech.}  \bvol{5},
  \pg{113--133}.

\bibitem[Batchelor \& Townsend(1956)]{Batchelor1956}
{\sc \au{Batchelor, G.~K.} \& \au{Townsend, A.A.}} \yr{1956}  \at{Turbulent
  diffusion}.  \jt{Surveys in Mechanics, Cambridge University Press}  \pg{pp.
  352--399}.

\bibitem[Benjamin(1968)]{Ben1968}
{\sc \au{Benjamin, T.B.}} \yr{1968}  \at{Gravity currents and related
  phenomena}.  \jt{Journal of Fluid Mechanics}  \bvol{31},  \pg{209--248}.

\bibitem[Cantero {\em et~al.\/}(2007)Cantero, Lee, Balachandar \&
  Garcia]{Cantero2007}
{\sc \au{Cantero, M.I.}, \au{Lee, J.R.}, \au{Balachandar, S.} \& \au{Garcia,
  M.H.}} \yr{2007}  \at{On the front velocity of gravity currents}.
  \jt{Journal of Fluid Mechanics}  \bvol{586},  \pg{1--39}.

\bibitem[Chen {\em et~al.\/}(2014)Chen, Zhong, Wang \& Li]{Chen2014}
{\sc \au{Chen, Q.}, \au{Zhong, Q.}, \au{Wang, X.} \& \au{Li, D.}} \yr{2014}
  \at{An improved swirling-strength criterion for identifying spanwise vortices
  in wall turbulence}.  \jt{Journal of Turbulence}  \bvol{15}~(2),
  \pg{71--87}.

\bibitem[Crimaldi(2008)]{Crimaldi}
{\sc \au{Crimaldi, J.}} \yr{2008}  \at{Planar laser induced fluorescence in
  aqueous flows}.  \jt{Experiments in Fluids}  \bvol{44}~(6),  \pg{851--863}.

\bibitem[Dimotakis(2000)]{Dimotakis2005}
{\sc \au{Dimotakis, P.E.}} \yr{2000}  \at{The mixing transition in turbulent
  flows}.  \jt{Journal of Fluid Mechanics}  \bvol{409},  \pg{69--98}.

\bibitem[Eckart(1948)]{Eckart1948}
{\sc \au{Eckart, C.}} \yr{1948}  \at{An analysis of the stirring and mixing
  processes in incompressible fluids}.  \jt{J. Mar. Res.}  \bvol{7},
  \pg{265--275}.

\bibitem[Ellision \& Turner(1959)]{ET}
{\sc \au{Ellision, T.H.} \& \au{Turner, J.S.}} \yr{1959}  \at{Turbulent
  entrainment in stratified flows}.  \jt{Journal of Fluid Mechanics}
  \bvol{6}~(03),  \pg{423--448}.

\bibitem[Fernando(1991)]{Fernando}
{\sc \au{Fernando, H.J.S.}} \yr{1991}  \at{Turbulent mixing in stratified
  flows}.  \jt{Journal of Fluid Mechanics}  \bvol{23},  \pg{455--493}.

\bibitem[Hannoun {\em et~al.\/}(1988)Hannoun, Fernando \& E.J.List]{Han1988}
{\sc \au{Hannoun, I.A.}, \au{Fernando, H.J.S.} \& \au{E.J.List}} \yr{1988}
  \at{Turbulence structure near a sharp density interface}.  \jt{Journal of
  Fluid Mechanics}  \bvol{189},  \pg{189--209}.

\bibitem[Hinze(1975)]{Hinze1975}
{\sc \au{Hinze, J.O.}} \yr{1975}  \at{Turbulence}.  \jt{McGraw-Hill College}
  \bvol{2nd edition},  \pg{790}.

\bibitem[Hunt {\em et~al.\/}(1983)Hunt, Rottman \& Britter]{Hunt1983}
{\sc \au{Hunt, J.C.R.}, \au{Rottman, J.W.} \& \au{Britter, R.E.}} \yr{1983}
  \at{Some physical processes involved in the dispersion of dense gases}.
  \jt{In Proc. IUTAM Symp. On Atmospheric Dispersion of Heavy Gases and Small
  Particles}  \pg{pp. 361--395}.

\bibitem[Huppert \& Simpson(1980)]{Huppert1980}
{\sc \au{Huppert, H.} \& \au{Simpson, J.~E}} \yr{1980}  \at{The slumping of
  gravity currents}.  \jt{Journal of Fluid Mechanics}  \bvol{99},
  \pg{785--799}.

\bibitem[Keulegan(1957)]{Keulegan1957}
{\sc \au{Keulegan, G.H.}} \yr{1957}  \at{An experimental study of the motion of
  saline water from locks into fresh water channels}.  \jt{US National Bur.
  Stand. Rep} ~(5168).

\bibitem[Klemp {\em et~al.\/}(1994)Klemp, Rotunno \& Skamarock]{Klemp1994}
{\sc \au{Klemp, J.B.}, \au{Rotunno, R.} \& \au{Skamarock, W.C.}} \yr{1994}
  \at{On the dynamics of gravity currents in a channel}.  \jt{Journal of Fluid
  Mechanics}  \bvol{269},  \pg{169--198}.

\bibitem[Koochesfahani \& Dimotakis(1986)]{Dimotakis1986}
{\sc \au{Koochesfahani, M.M.} \& \au{Dimotakis, P.E.}} \yr{1986}  \at{Mixing
  and chemical reactions in a turbulent liquid mixing layer}.  \jt{Journal of
  Fluid Mechanics}  \bvol{170},  \pg{83--112}.

\bibitem[Koop \& Browand(1979)]{Koop1979}
{\sc \au{Koop, C.G.} \& \au{Browand, F.K.}} \yr{1979}  \at{Instability and
  turbulence in a stratified fluid with shear}.  \jt{Journal of Fluid
  Mechanics}  \bvol{93},  \pg{135--139}.

\bibitem[Linden(1979)]{Linden1979}
{\sc \au{Linden, P.F.}} \yr{1979}  \at{Mixing in stratified fluids}.
  \jt{Geophysical and Astrophysical Fluid Dynamics}  \bvol{12},  \pg{3--23}.

\bibitem[Manins(1976)]{Manins}
{\sc \au{Manins, P.C}} \yr{1976}  \at{Intrusions into a stratified fluid}.
  \jt{Journal of Fluid Mechanics}  \bvol{74},  \pg{547--560}.

\bibitem[McC.Hogg \& Ivey(2003)]{Ivey2003}
{\sc \au{McC.Hogg, A.} \& \au{Ivey, G.N.}} \yr{2003}  \at{The kelvin-helmholtz
  to holmboe instability transition in stratified exchange flows}.  \jt{Journal
  of Fluid Mechanics}  \bvol{477},  \pg{339--362}.

\bibitem[Mirajkar {\em et~al.\/}(2015)Mirajkar, Tirodkar \&
  Balasubramanian]{Harish2015}
{\sc \au{Mirajkar, Harish~N}, \au{Tirodkar, Siddhesh} \& \au{Balasubramanian,
  Sridhar}} \yr{2015}  \at{Experimental study on growth and spread of dispersed
  particle-laden plume in a linearly stratified environment}.  \jt{Environmenal
  Fluid Mechanics}  \bvol{15}~(6),  \pg{1241--1262}.

\bibitem[Morton {\em et~al.\/}(1956)Morton, Taylor \& Turner]{MTT}
{\sc \au{Morton, B.R.}, \au{Taylor, G.I.} \& \au{Turner, J.S.}} \yr{1956}
  \at{Turbulent gravitational convection from maintained and instantaneous
  sources}.  \jt{The Royal Society Proc. A}  \bvol{234}~(1196),  \pg{1--23}.

\bibitem[Ottolenghi {\em et~al.\/}(2016)Ottolenghi, Adduce, Inghilesi, Armenio
  \& Roman]{Ottolenghi2016}
{\sc \au{Ottolenghi, L.}, \au{Adduce, C.}, \au{Inghilesi, R.}, \au{Armenio, V.}
  \& \au{Roman, F.}} \yr{2016}  \at{Entrainment and mixing in unsteady gravity
  currents}.  \jt{Journal of Hydraulic Research}  \bvol{54}~(5),
  \pg{541--557}.

\bibitem[Princevac {\em et~al.\/}(2005)Princevac, Fernando \&
  Whiteman]{Princevac2005}
{\sc \au{Princevac, M.}, \au{Fernando, H.J.S.} \& \au{Whiteman, C.D.}}
  \yr{2005}  \at{Turbulent entrainment into natural gravity-driven flows}.
  \jt{Journal of Fluid Mechanics}  \bvol{533},  \pg{259--268}.

\bibitem[Scarano(2002)]{Scarano}
{\sc \au{Scarano, F.}} \yr{2002}  \at{Iterative image deformation methods in
  piv}.  \jt{Measurement Science and Technology}  \bvol{13}~(1),  \pg{R1}.

\bibitem[Shin {\em et~al.\/}(2004)Shin, Dalziel \& Linden]{Shin2004}
{\sc \au{Shin, J.O.}, \au{Dalziel, S.B.} \& \au{Linden, P.F.}} \yr{2004}
  \at{Gravity currents produced by lock exchange}.  \jt{Journal of Fluid
  Mechanics}  \bvol{521},  \pg{1--34}.

\bibitem[Simpson(1982)]{EJS1982}
{\sc \au{Simpson, EJ}} \yr{1982}  \at{Gravity currents in the laboratory,
  atmosphere, and ocean}.  \jt{Annual Review of Fluid Mechanics}  \bvol{14},
  \pg{213--234}.

\bibitem[Simpson \& Britter(1979)]{Simpson1979b}
{\sc \au{Simpson, J.E.} \& \au{Britter, R.E.}} \yr{1979}  \at{The form of the
  head of an intrusive gravity currents}.  \jt{Geophysical Journal of Research
  Astron. Soc}  \bvol{57},  \pg{289}.

\bibitem[Smyth \& Peltier(1991)]{Smyth1991}
{\sc \au{Smyth, W.D.} \& \au{Peltier, W.R.}} \yr{1991}  \at{Instabilty and
  transition in finite-amplitude kelvin--helmholtz instability and transition
  in finite-amplitude kelvin-helmholtz and holmboe waves}.  \jt{Journal of
  Fluid Mechanics}  \bvol{228},  \pg{387--415}.

\bibitem[Strang \& Fernando(2001a)]{Strang2001a}
{\sc \au{Strang, E.J.} \& \au{Fernando, H.J.S.}} \yr{2001a}  \at{Entrainment
  and mixing in stratified shear flows}.  \jt{Journal of Fluid Mechanics}
  \bvol{428},  \pg{349--386}.

\bibitem[Strang \& Fernando(2001b)]{Strang2001b}
{\sc \au{Strang, E.J.} \& \au{Fernando, H.J.S.}} \yr{2001b}  \at{Vertical
  mixing and transports through a stratified shear layer}.  \jt{J. Physical
  Oceanography}  \bvol{31},  \pg{2026--2048}.

\bibitem[Thorpe(1973)]{Thorpe1972}
{\sc \au{Thorpe, S.A.}} \yr{1973}  \at{Experiments on instability and
  turbulence in a stratified shear flow}.  \jt{Journal of Fluid Mechanics}
  \bvol{61},  \pg{731--751}.

\bibitem[Townsend(1976)]{Townsend}
{\sc \au{Townsend, A.A.}} \yr{1976}  \at{The structure of turbulent shear
  flow}.  \jt{Cambridge University Press}  \bvol{2nd edition},  \pg{127}.

\bibitem[Turner(1986)]{Turner1986}
{\sc \au{Turner, J.S.}} \yr{1986}  \at{Turbulent entrainment: the development
  of the entrainment assumptions, and its application to geophysical flows}.
  \jt{Journal of Fluid Mechanics}  \bvol{173},  \pg{431--471}.

\bibitem[Xu \& Chen(2012)]{Xu2012}
{\sc \au{Xu, D.} \& \au{Chen, J.}} \yr{2012}  \at{Experimental study of
  stratified jet by simultaneous measurements of velocity and density fields}.
  \jt{Experiments in Fluids}  \bvol{53}~(1),  \pg{145--162}.

\bibitem[Yih(1965)]{Yih1965}
{\sc \au{Yih, C.S.}} \yr{1965}  \at{Dynamics of nonhomogenous fluids}.  \jt{New
  York: Macmillan} .

\bibitem[Zhong {\em et~al.\/}(2015)Zhong, Li, Chen \& Wang]{Zhong2015}
{\sc \au{Zhong, Q.}, \au{Li, D.}, \au{Chen, Q.} \& \au{Wang, X.}} \yr{2015}
  \at{Coherent structures and their interactions in smooth open channel flows}.
   \jt{Environmenal Fluid Mechanics}  \bvol{15}~(3),  \pg{653--672}.

\end{thebibliography}
\end{document}